\documentclass[12pt]{article}
\usepackage{graphicx,amssymb,epsfig,epsf} 
 \usepackage{ulem}
\usepackage[usenames]{color}
\usepackage{rotating}
\usepackage{epsfig}
\usepackage{slashed}
\usepackage{amsmath}
\usepackage{pstricks}
\usepackage{color}

\begin{document}

\begin{flushright}
   { OSU-HEP-14-10,\\
   NSF-KITP-14-174
   }\\
\end{flushright}

\vskip 2pt

\begin{center}
{\large \bf Dark Matter, Parallel Universe and Multiple Higgs Signals at the ILC}\\

\vskip 20pt
{
Shreyashi Chakdar$^{a}$\footnote{chakdar@okstate.edu}, {Kirtiman Ghosh$^{a}$\footnote{kirti.gh@gmail.com}}},
 S. Nandi$^{a}$\footnote{s.nandi@okstate.edu} 
   \\
\vskip 10pt
{$^{a1, a2, a3}$Department of Physics and Oklahoma Center for High Energy Physics,\\
Oklahoma State University, Stillwater, OK 74078-3072, USA.}\\
\vskip 10pt
{ $^{a1}$ Kavli Institute for Theoretical Physics,\\
University of California, Santa Barbara, 
CA 93106, USA.}

\end{center}

\vskip 5pt
\abstract
{
\noindent The existence of dark matter is now well established by several indirect experiments. Several candidates for dark matter
has also been proposed. However, the dark matter can just be like our ordinary matter in a parallel Universe with both Universes being
described by their own non-abelian gauge symmetries forbidding any kinetic mixing. However, the quartic Higgs interactions involving Higgs fields between the two Universes are allowed by the symmetries of the model.  The ensuing mixing between the two lightest Standard Model like Higgses gives rise to interesting signatures at the proposed international electron-positron collider (ILC) specially in the case when mass splitting between the two surviving light Higgs bosons are small ($\sim$ 100 MeV) so that they can not be resolved at the LHC.

}
\vskip 5pt
\vskip 40pt
\section{Introduction}

The Standard Model (SM) of particle physics based on the gauge symmetry $SU(3)_C \times SU(2)_L \times U_Y(1)$ contains our best formulation to date for understanding the observed classification of elementary particles and their interactions upto energies of about 1 TeV.
The recent discovery of the Higgs boson  \cite{ATLAS_higgs} completed the framework of SM, but there are still  many unanswered theoretical questions as well as many unexplained experimental phenomena. For example, the charge quantization of the elementary particles can not be explained in the framework of the SM gauge symmetry. In order to explain the charge quantization, one needs to enlarge the SM gauge symmetry to $SU(4)_C \times SU(2)_L \times SU(2)_R$ 
with the lepton number as the fourth color or grand unifying all three interaction in $SU(5)$ \cite{GG} or $SO(10)$ {GUT} \cite{gfm}.
Similarly, existence of the observed  non-zero neutrino masses can not be explained in the SM. Another major drawback of the SM is the  absence of any dark matter(DM) candidate. In recent years, an ever-increasing number of measurements have been performed and the acquired data  support strongly the existence of the dark matter in our Universe. For example, the mass-energy densities of the universe computed from the Wilkinson Microwave Anisotropy Probe (WMAP) \cite{WMAP1}data shows that the present day dynamics of the Universe is driven essentially by the dark energy (DE). The major fraction of the total energy of our universe is the dark energy and is currently estimated to be 68.3\%. The  ordinary baryonic matter, i.e. nuclei and electrons, constitute about 4.9\% while the remaining 26.8\% is the dark matter. Thus dark matter in the universe is about five times larger than the visible matter. The SM of particle physics explains the properties and interactions of the visible matter but it does not include any  dark matter candidate. This provide strong motivation beyond the SM (BSM) scenarios. Many extensions of the SM model offer candidates for dark matter such as  new weakly interacting massive particles (WIMPs) \cite{dm}, axions \cite{ww}, the lightest stable particles in supersymmetry \cite{dm}, or the lightest Kaluza-Klein particle in extra dimensions \cite{kkdm}. \\

Another possible scenario for the dark matter is the presence of new dark particles not charged under SM gauge symmetry which are analogous to the ordinary particles belonging to a parallel universe. Such a parallel universe  naturally appears in the superstring theory with the $E_8 \times E'_8$ gauge symmetry before compactification \cite{chsw}. In a recent work \cite{plbus}, we assumed the dark matter to be just like the ordinary matter in a parallel universe. The dark protons, dark electrons and the corresponding dark nucleis belonging to the parallel universe constitute the dark matter in this case. The parallel universe scenario is motivated to explain the measured dark matter density of the universe which is five times larger than the ordinary matter density and this is achieved by assuming the QCD scale in the dark sector ($\Lambda_{DS}$) to be five times larger than the QCD scale in the visible sector ($\Lambda_{VS}$). Any gauge or Yukawa interactions  between the two universes are prohibited  by postulating that each of these universes is described  by separate non-abelian gauge symmetries. However, the two sectors can still interact via the Higgs bosons of respective sectors. If the mass splitting between this two Higgs bosons are $\sim 100$ MeV, they can not be resolved as separate mass-peaks at the LHC, but can be accesible as separate peaks in the Electron-Positron colliders such as the proposed ILC \cite{Djouadi:2007ik}. In this work, we look into the Electron-Positron Collider Phenomenology in case of $\sim$ 100 MeV mass splittings between the two respective lightest SM like  Higgses of our universe and the parallel universe.

\section{Model and Formalism}

  The gauge symmetry we use is  $ SU(4)_C \times SU(2)_L \times SU(2)_R $ for our universe and $ SU(4)'_C \times SU(2)'_L \times SU(2)'_R $ for the parallel universe. It is important to note that the choice of this non-abelian symmetry not only explains charge quantization (as in Pati-Salam model \cite{PS}), but also avoids the kinetic mixing between the ordinary photon ($\gamma$) and parallel photon ($\gamma '$). The 21 gauge bosons belong to the adjoint representations $(15,1,1)$, $(1,3,1)$ and $(1,1,3)$. $(15,1,1)$ contain the 8 usual colored gluons, 6 lepto-quark gauge bosons $(X, \bar{X})$ and one $(B-L)$ gauge boson \cite{mm}. $(1,3,1)$ contain the 3 left handed weak gauge bosons, while $(1,1,3)$ contain the 3 right handed weak gauge bosons. The parallel universe contains the corresponding parallel gauge bosons. As far as the electro-weak gauge interactions are concerned, we assume an exact symmetry between our universe and the parallel universe. However, we have assumed different couplings for $SU(4)$ and $SU(4)'$ interactions in order to explain the dark matter  to be five times larger than visible matter. Therefore, the model has 3 gauge coupling constants: $g_4$ for $SU(4)$ color which we identify with the strong coupling constant ($g_s^{VS}$) for our universe (namely, the visible sector (VS); $g'_4$ for $SU(4)'$ color which we identify with the strong coupling constant ($g_s^{DS}$) for the parallel universe (namely the dark sector (DS)). The electroweak coupling for our universe is denoted by  $g$. The corresponding electroweak couplings for the parallel universe is denoted by  g'. ($g_L = g_R=g $ and $ g'_L = g'_R=g'$) by left-right symmetry. We assume that $g = g'$. The fermions belong to the fundamental representations $( 4, 2,1) + (4,1,2)$. The 4 represent three colors of quarks and the lepton as the 4th color, whereas $(2,1)$ and $(1,2)$ represent the left and right handed doublets.

$SU(4)$ color symmetry is spontaneously broken to $SU(3)_C \times U(1)_{B-L}$ in the usual Pati-Salam way using  the Higgs fields $(15, 1,1)$ at a scale $V_c$. The limit on the scale of this symmetry breaking $V_c$ comes from the upper limit of the rare decay mode $K_L \rightarrow \mu e$ \cite{Ritchie:1993ua}.
 $SU(2)_L \times SU(2)_R \times U(1)_{B-L}$ can be broken to the SM symmetry using the Higgs representations  $(1,2,1)$ and {(}$1,1,2)$ at a scale $V_{LR}$. Similar Higgs representations are used to break the symmetry in the dark sector. A study of the Higgs potential shows that there exists a parameter space where only one neutral Higgs in the bi-doublet remains light, and becomes very similar to the SM Higgs in our universe \cite{Senjanovic}. All other Higgs fields become very heavy compared to the EW scale. Similar is true in the parallel universe. 

In this model, interaction between fermions and/or gauge bosons of dark sector and visible sector are forbidden by the gauge symmetry. However, quartic Higgs interactions of the form $\lambda (H_{VS}^{\dag}H_{VS})(H_{DS}^{\dag}H_{DS})$ (where $H_{VS}$ and $H_{DS}$ symbols denote the Higgs fields in the visible sector and dark sector respectively) are allowed and give rise to mixing between the Higgses of dark and visible sector. The mixing between the lightest Higgses of dark sector (denoted by $h_2$) and visible sector (denoted by $h_1$) gives rise to interesting phenomenological implications at the collider experiments and hence make this model testable at the collider experiments. We denote the two mass eigenstates to be $h_{1}^{(p)}$ and $h_{2}^{(p)}$. Identifying one of the light mass eigenstates with as the Higgs boson (denoted by $h_{SM}$) observed by the Large Hadron Collider (LHC), in Ref.~\cite{plbus}, we have shown that the LHC data can not rule out the existence of the other mass eigenstate (namely, the dark Higgs) when the mass difference between these two lighest Higgs bosons
 is less than 2 GeV. In this work, we have discussed the signatures of these Higgs bosonss when their mass differences are $40$ and $500$ MeV in the context of proposed ILC  with the initial center-of-mass energy $\sim 250$ GeV.   

\begin{figure}
\begin{center}
\includegraphics[width=8.5 cm,height=10.5cm,angle=-90]{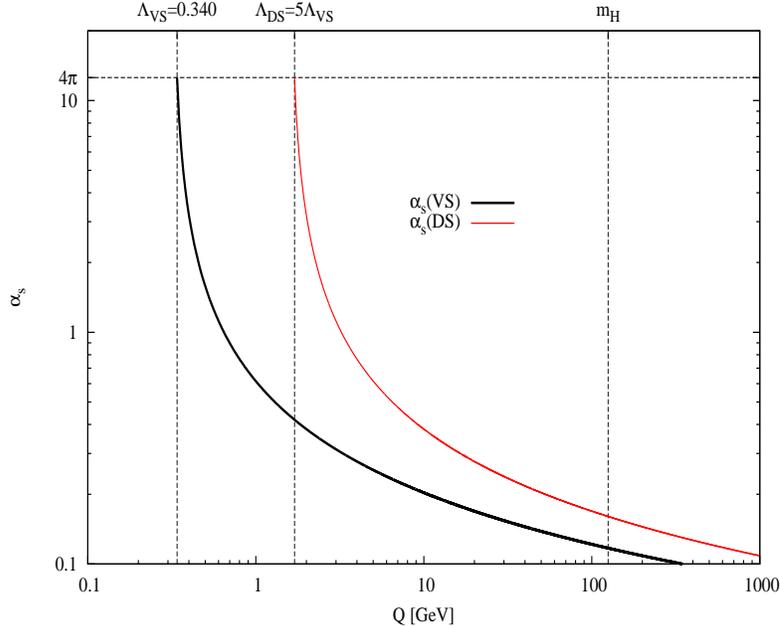}
\end{center}
\caption{Running of the strong coupling constant in the visible sector ($\alpha_S^{VS}(Q)$) and dark sector ($\alpha_S^{DS}(Q)$)}
\label{running}
\end{figure}

The main motivation for postulating this kind of parallel universe scenario is to explain the dark matter density which is five time larger than the ordinary matter density. The particles analogous to the proton and neutron in the dark sector, namely, the dark proton and dark neutron, are stable due to the conservation of the dark baryon number. Moreover, the gauge symmetry and the particle content  of the model does not allow any gauge or Yukawa interactions of dark protons and dark neutrons with the visible sector particles. Therefore, in the framework of this model, the  dark protons and dark neutrons are  the candidates for the dark matter. If the dark protons and dark neutrons in the parallel universe are about five times heavier than the protons and neutrons of our universe, then that will naturally explain why the dark matter of the universe is about five times the ordinary matter. The Lagrangian quark masses for the up and down quarks are only of order of 10 MeV or less. Therefore,$99\%$ of the mass of the proton or neutron arises from the strong interaction of the constituent quarks and the gluons. The mass scale associated with these interactions is set by the value of the three-quark QCD scale $\Lambda_{QCD}$. Therefore, one can easily achieve a dark proton or neutron mass which is five times larger than the visible proton or neutron mass by assuming the QCD scale in the dark sector ($\Lambda_{DS}$) to be five times larger than the QCD scale in the visible sector ($\Lambda_{VS}=340$ MeV \cite{pdg_qcd}). Different QCD scales give rise to different running of the strong coupling constant in the visible sector ($\alpha_S^{VS}(Q)$) and dark sector ($\alpha_S^{DS}(Q)$). In Fig.~\ref{running}, we have presented the running of the strong coupling constant in the visible sector and dark sector. Fig.~\ref{running} shows that $\alpha_S^{DS}(Q=m_H=125~{\rm GeV})$= $ 1.4~ \alpha_S^{VS}(Q=m_H=125~{\rm GeV})$.

Before going into the details of the collider phenomenology, it is important to discuss the masses and mixing of the physical light Higgs states resulting from the bi-linear terms involving the lightest visible sector and dark sector  Higgses,
\begin{equation}
{\cal L}_{scalar} \supset m_{VS}^2h_1^2 + m_{DS}^2h_2^2 + 2 \lambda v_{VS} v_{DS} h_1 h_2
\end{equation} 
where, $v_{VS}$ and $v_{DS}$ are the electroweak symmetric breaking scale in the visible sector and dark sector respectively. As a consequence of the  symmetry of the Higgs fields in   the EW sector between our universe and the parallel universe, the two  electroweak VEV's are same, $v_{VS}=v_{DS}=v_{SM}=250$ GeV. $m_{VS}$, $m_{DS}$ and $\lambda$ are the free parameters in the theory and the masses ($m_{h_1^{(p)}}$ and $m_{h_2^{(p)}}$) and mixing between physical light Higgs states (denoted by $h_1^{(p)}$ and $h_{2}^{(p)}$) are determined by these parameters:   
\begin{eqnarray}
h_{1}^{(p)}&=&{\rm cos}\theta ~h_1 + {\rm sin}\theta ~h_2, \nonumber\\
h_{2}^{(p)}&=&-{\rm sin}\theta ~h_1 + {\rm cos}\theta ~h_2,
\end{eqnarray}
where the masses and the mixing angle of these physical states are given by,
\begin{eqnarray}
m_{h_1^{(p)},h_2^{(p)}}^2&=&\frac{1}{2}[(m_{VS}^2+m_{DS}^2)\mp\sqrt{(m_{VS}^2-m_{DS}^2)^2+4\lambda^2v_{VS}^2v_{DS}^2}], \nonumber\\
{\rm tan}2\theta &=&\frac{2\lambda ~v_{VS}~v_{DS}}{m_{DS}^2-m_{VS}^2}.
\end{eqnarray}
In the framework of this model, we have two light physical neutral Higgs ($h_1^{(p)}$ and $h_{2}^{(p)}$) states. Out of these two Higgs states, we define the SM like Higgs.
if ${\rm cos}\theta > {\rm sin}\theta$ then $h_{1}^{(p)}$ is like the SM Higgs; whereas ${\rm cos}\theta < {\rm sin}\theta$, the$h_{2}^{(p)}$ is dominantly the SM Higgs. We will consider the phenomenology for the case of $h_{1}^{(p)}$ is dominantly the SM Higgs. 

The visible sector light Higgs weak eigenstate $h_1$ interacts only with the visible sector fermions ($f$) and gauge bosons ($V$), whereas the dark sector light Higgs weak eigenstate $h_2$ interacts only with the dark fermions $f_D$ and dark gauge bosons $V_D$. However, due to the mixing, the physical light Higgses interact with both the visible particles and dark particles and thus, both of them can be produced at the collider experiments. Moreover, both the mass eigenstates can decay into a pair of visible or dark particle. Dark particles do not have any interactions with the visible particles and thus they remain invisible in the collider experiments. As a result, the decay of the Higgs bosons into the dark particles give rise to invisible Higgs signature at the collider experiments.  The coupling of the physical states $h_1^{(p)}$ and $h_2^{(p)}$ with the visible as well as dark fermions and gauge bosons can be written as a product of corresponding SM coupling and sine or cosine of the mixing angle. Therefore, the decay widths and hence, the decay branching ratios of the physical states $h_1^{(p)}$ and $h_2^{(p)}$ into pair of visible as well as dark particles can be computed in terms of the SM Higgs decay widths/branching ratios and the mixing angle. For example, the decay widths of $h_{1}^{(p)} (h_{2}^{(p)})$ into a pair of fermions of the visible and dark sector are given by $\Gamma_{SM}^{H\to f\bar f} {\rm cos}^2\theta$ ($\Gamma_{SM}^{H\to f\bar f} {\rm sin}^2\theta$) and $\Gamma_{SM}^{H\to f\bar f} {\rm sin}^2\theta$ ($\Gamma_{SM}^{H\to f\bar f} {\rm cos}^2\theta$) respectively, where $\Gamma_{SM}^{H\to f\bar f}$ is the decay width of the SM Higgs into $f\bar f$. Except for the dark gluon, similar relations hold for the other visible ($Z$-boson, $W$-boson, photon and gluon) and dark ($Z_D$-boson, $W_{D}$-boson and dark photon) gauge bosons. The QCD coupling in the dark sector is about $k=1.4$ times larger than the QCD coupling in the visible sector. Therefore, the Higgs coupling with dark gluon in this model is enhanced by a factor about $k=1.4$. The visible and invisible branching ratios for the physical Higgs states $h_1^{(p)}$ and $h_2^{(p)}$ are given by,
\begin{eqnarray}
{\mathcal Br}[h^{(p)}_1 (h^{(p)}_2)\to x^i x^i] &=& \frac{{\mathcal Br}[H_{SM}\to x^i x^i]{\rm cos}^2\theta ({\rm sin}^2\theta)}{1+(k^2-1){\mathcal Br}[H_{SM}\to gg]{\rm sin}^2\theta ({\rm cos}^2\theta)},\nonumber\\
{\mathcal Br}[h^{(p)}_1(h^{(p)}_2)\to \sum_i x^i_D x^i_D]&=&\frac{1+({k}^2-1){\mathcal Br}[H_{SM}\to gg]}{1+(k^2-1){\mathcal Br}[H_{SM}\to gg]{\rm sin}^2\theta({\rm cos}^2\theta)}{\rm sin}^2\theta({\rm cos}^2\theta), 
\end{eqnarray}
where, $x^i$ and $x^i_D$ represent the visible and dark particle, respectively, and ${\mathcal Br}[H_{SM}\to x^i x^i]$ corresponds to the branching ratio of the SM Higgs into $x^i x^i$.

\section{Collider Phenomenology}

After discussing the mixings and the decays of the physical Higgs states $h_1^{(p)}$ and $h_2^{(p)}$, we are now equipped
enough to discuss the collider phenomenology in the context of an electron-positron collider. In this work, we are particularly interested in the scenario in which two light  physical Higgs states are quasi-degenerate. Due to the large background, it will be challenging to distinguish such a scenario at the hadron collider experiments. The advantages of an electron-positron collider compared to a hadron collider are the cleanliness of the environment, the precision of the measurements and the large number of Higgs bosons production. Therefore, it could be possible for an  electron-positron collider to probe a scenario with two quasi-degenerate Higgs bosons. It has recently been shown by the {\it   International Large Detector (ILD) Concept Group} in Ref.~\cite{ILD_resolution} that the proposed electron-positron collider will be able to determine the Higgs boson mass with a statistical precision of 40 MeV. Motivated by the results of Ref.~\cite{ILD_resolution}, in our analysis, we have considered two different mass splittings between the Higgs bosons:
\begin{itemize}
\item {\bf Scenario I:} We have considered the mass splitting between the two Higgs bosons to be about 40 MeV. Therefore, electron-positron collider can not resolve two Higgs bosons mass peaks in this case. The LHC experiment has already observed a Higgs boson with mass about 125 GeV. Therefore, we have assumed one Higgs boson mass to be 124.98 GeV and the other Higgs boson mass is 125.02 GeV.
\item {\bf Scenario II:} In this case, we assume relatively large mass splitting (about 500 MeV) between the two Higgs bosons so that the electron-positron collider can resolve the Higgs bosons mass peaks. The numerical values of the Higgs boson masses are chosen to be  124.75 GeV and 125.25 GeV.  
\end{itemize}

\begin{figure}
\begin{center}
\includegraphics[width=6.5 cm,height=18cm,angle=-90]{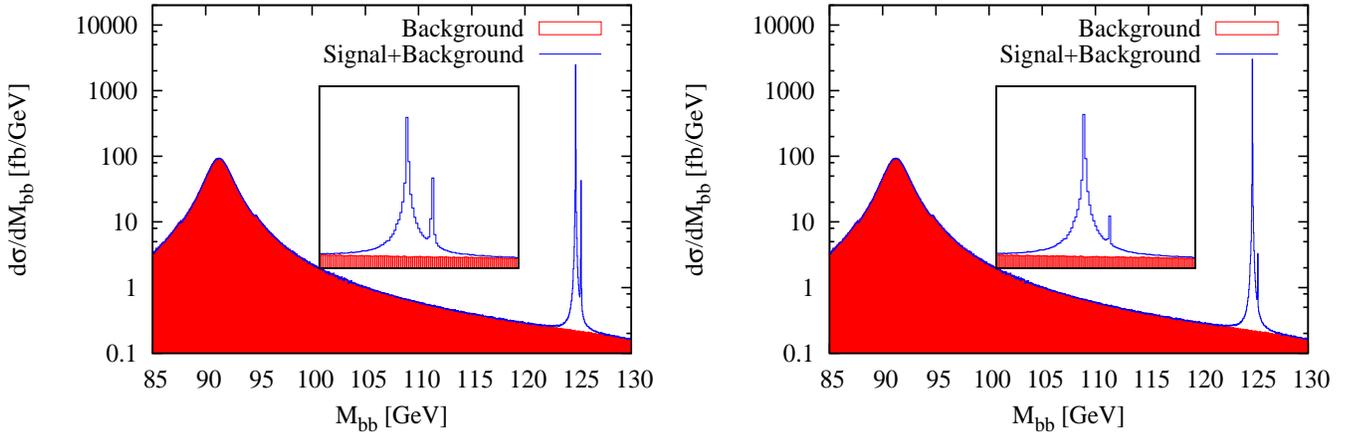}
\end{center}
\caption{Invariant mass distributions of $b\bar b$ pairs for two different values of the mixing angle, $\theta=20^0$ (left panel) and $10^0$ (right panel). The masses of the two physical Higgs states are given by $m_{h_{SM}}=124.75$ GeV and $m_{h_{DH}}=125.25$ GeV. $123~{\rm GeV}<m_{bb}<127~{\rm GeV}$ regions are magnified in the insets.}
\label{inv_mass_large}
\end{figure}    

At the electron-positron collider the main production mechanism of the Higgs boson is the Higgs-strahlung process i.e., the production of the Higgs boson in association with a $Z$-boson: $e^+e^- \to ZH$. The Higgs-strahlung process is an s-channel process so that its production cross section is maximal just above the threshold of the process. As a result, for the proposed electron-positron collider to study the observed Higgs boson properties in detail, we should start with the $e^+e^-$ collision initial center-of-mass energy of 250 GeV at which the Higgs-strahlung cross-section reaches its  maximum value for a $125$ GeV  Higgs boson mass. In our analysis, we have also considered 250 GeV center-of-mass energy for the electron-positron collider. In the framework of the present model, both physical Higgs states will be produced in association with a $Z$-boson. The production cross-sections of $h_1^{(p)}$ and $h_2^{(p)}$ are given in terms of the SM Higgs production cross-section ($\sigma_{SM}(ZH)$) and the mixing angle: $\sigma(e^+e^- \to Z h_1^{(p)})={\rm cos}^2\theta \sigma_{SM}(ZH)$ and $\sigma(e^+e^- \to Z h_2^{(p)})={\rm sin}^2\theta \sigma_{SM}(ZH)$ . From the Higgs-strahlung process, the Higgs signature could be detected and hence, Higgs mass could be measured by the direct Higgs decays and the recoiling to the $Z$-boson. In our analysis, we have studied both the direct Higgs decays as well as  the recoiling to the $Z$-boson. The recoil mass to the $Z$-boson is the invariant mass of the decay products against which the $Z$-boson recoils assuming the collision occurs at the nominal center of mass energy $\sqrt s$ and is defined as,
$$
M_{recoil}^2=(\sqrt s-E_Z)^2-|\vec p_Z|^2=s+M_Z^2-2E_Z\sqrt s,
$$
where $M_Z$ denote the mass of the $Z$-boson as reconstructed from the decay products of the $Z$-boson and $E_Z$ is the corresponding energy. Recoil mass to the $Z$-boson do not depend on the decay products of the Higgs boson and hence, provides an unique opportunity for the reconstruction of the Higgs mass from its invisible decays. In the framework of the present model, both  $h_1^{(p)}$ and $h_2^{(p)}$ decay to the dark sector particles which remain invisible in the detector. Therefore, in order to detect the decays $h_1^{(p)}$ and $h_2^{(p)}$ into a pair of dark particles, we have used the recoil mass to the $Z$-boson.

\begin{figure}
\begin{center}
\includegraphics[width=6.5 cm,height=18cm,angle=-90]{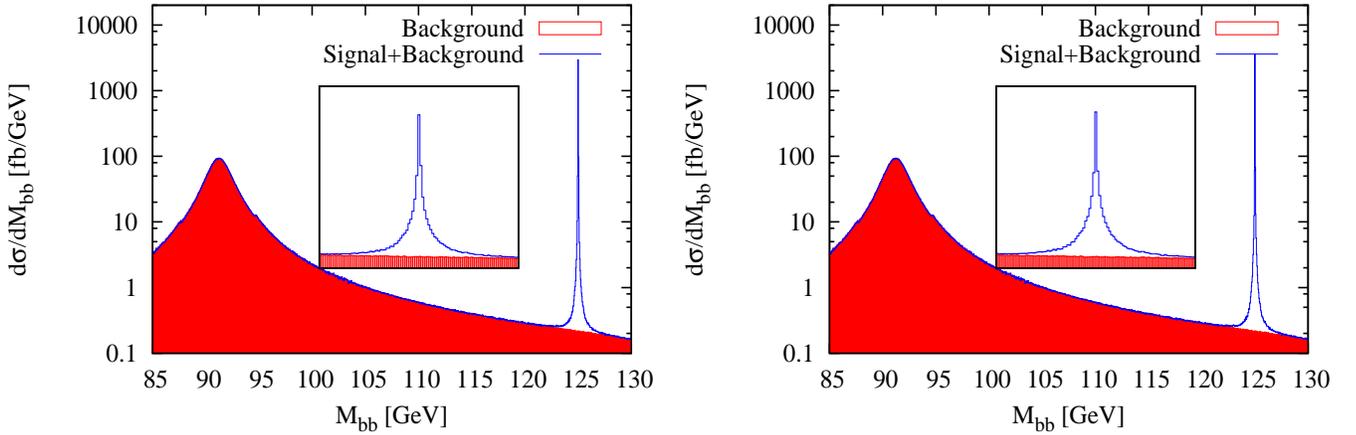}
\end{center}
\caption{Same as Fig.~\ref{inv_mass_large} for $m_{h_{SM}}=124.98$ GeV and $m_{h_{DH}}=125.02$ GeV.}
\label{inv_mass_small}
\end{figure}    

We have used CalcHEP package \cite{calchep} for the simulation of the signal and the background. The recoil mass to the $Z$-boson  crucially depends on the initial state radiation (ISR) and beamstrahlung. CalcHEP implements the Jadach, Skrzypek and Ward expressions of Refs.~\cite{ISR} for the simulation of ISR. Whereas for the Beamstrahlung, we have used the parameterizations specified for the ILD project \cite{ILD}: {\bf beam size (x + y)} = 645:7 nm, {\bf bunch length} = 300 $\mu$m and {\bf bunch population} = $2\times 10^{10}$.

\begin{table}[h]

\begin{center}

\begin{tabular}{c|c|c}
\hline\hline

Kinematic Variable & Minimum value & Maximum value \\\hline\hline
$\Delta R(l^{+}l^{-})$ & 0.3  & -    \\
$\Delta R(bb)$ & 0.7  & -    \\
$\Delta R(l^\pm b)$ & 0.7  & -    \\
$p_{T}^{l^{+},l^{-}}$  & 10 GeV   & -    \\
$\eta_{l^{+},l^{-}}$ & -2.5  & 2.5 \\
$p_{T}^{b}$  & 20 GeV   & -    \\
$\eta_{b}$ & -2.5  & 2.5 \\
$M(l^{+}l^{-})$    & 80 GeV      & 100 GeV \\\hline\hline

\end{tabular}

\end{center}

\caption{Acceptance cuts,used in this calculation, on the kinematical variables for {\it
2-lepton + 2-b} signal.}

\label{cuts}
\end{table}

\begin{figure}
\begin{center}
\includegraphics[width=6.5 cm,height=18cm,angle=-90]{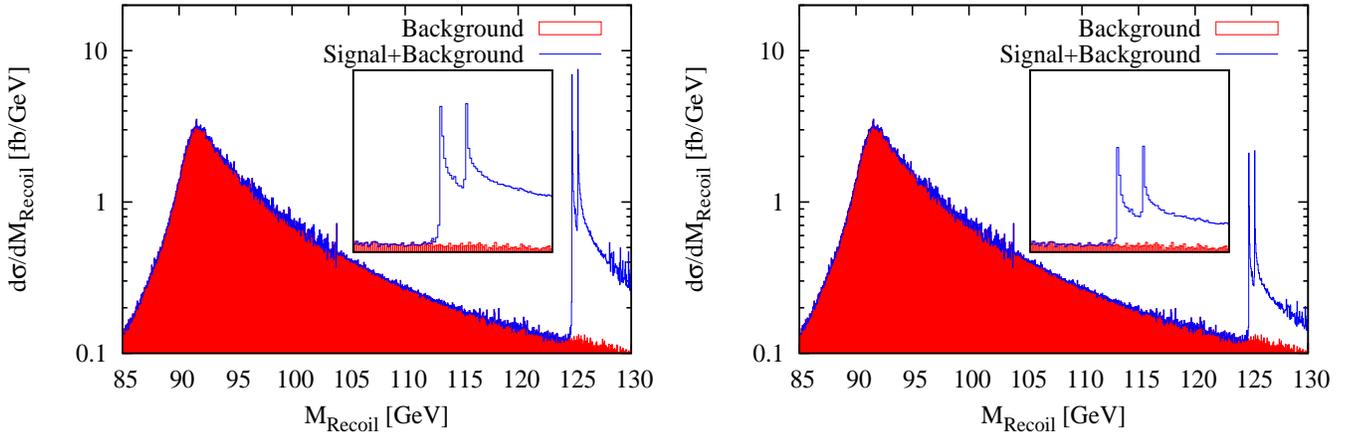}
\end{center}
\caption{Recoil mass distribution for invisible Higgs decays for two different values of the mixing angle, $\theta=20^0$ (left panel) and $10^0$ (right panel). The masses of the two physical Higgs states are given by $m_{h_{SM}}=124.75$ GeV and $m_{h_{DH}}=125.25$ GeV. $123~{\rm GeV}<m_{bb}<127~{\rm GeV}$ regions are magnified in the insets. }
\label{re_mass}
\end{figure}    

Higgs boson dominantly decays into a pair of bottom-quarks. For the reconstruction of the Higgs bosons from its decay products, we consider the decays of Higgs boson into a pair of $b$-quarks. For the associated $Z$-boson, we consider its leptonic (electron and muon) decay modes only. Therefore, the signal is characterized by two opposite sign same flavour leptons and two bottom quarks. The dominant background arises from the production of a pair of $Z$-bosons when one $Z$ decays in to a pair of b-quarks and another $Z$ decays leptonically. To parameterize detector acceptance and enhance signal to background ratio, we have imposed kinematic cuts, listed in Table~\ref{cuts}.  In Fig.~\ref{inv_mass_large}, we have presented the invariant mass distributions of $b\bar b$ pairs for two different values of the mixing angle, $\theta=20^0$ (left panel) and $10^0$ (right panel) for {\bf Scenario II}. We have used a bin size of 40 MeV. {\bf Scenario II} corresponds to relatively large mass splitting between the two Higgs bosons. As a result, in Fig.~\ref{inv_mass_large}, two characteristic mass peaks in the $b\bar b$ invariant mass distributions are clearly visible. Fig.~\ref{inv_mass_small} corresponds to the $b\bar b$ invariant mass distributions for the {\bf Scenario I}.

We have also studied the recoil mass to the $Z$-boson when the two Higgs bosons decay invisibly to a pair of dark sector particles. The $Z$-boson is assumed to decay into a pair of leptons (electrons and muons only). Although the branching ratio of the $Z \to l^+l^-$, where $l$ refers to $e$ or $\mu$, is only about 3.4\%, which is about 20 times smaller than that of the $Z \to q\bar q$, the high momentum resolution of leptons could overcome the shortage in statistics to gain even higher precision on the Higgs mass measurement. The signal in this case is characterized by two opposite sign same flavor leptons and missing energy. The dominant background in this case is again $ZZ$ production followed by leptonic decay of one $Z$-boson and invisible decay of the other $Z$-boson. The cuts listed in Table~\ref{cuts} are applied for the leptons. Fig.~\ref{re_mass} gives the background and signal+background recoil mass distributions for {\bf Scenario II} for two different values of the mixing angle, $\theta=20^0$ (left panel) and $10^0$ (right panel). The Higgs recoil mass distributions are crucially affected by the beamstrahlung and the initial state radiation which are responsible for the long tail of the distributions as visible in Fig.~\ref{re_mass}. However, 
the two Higgs bosons mass peaks are clearly visible in Fig.~\ref{re_mass}. 
\begin{figure}
\begin{center}
\includegraphics[width=8.5 cm,height=10.5cm,angle=-90]{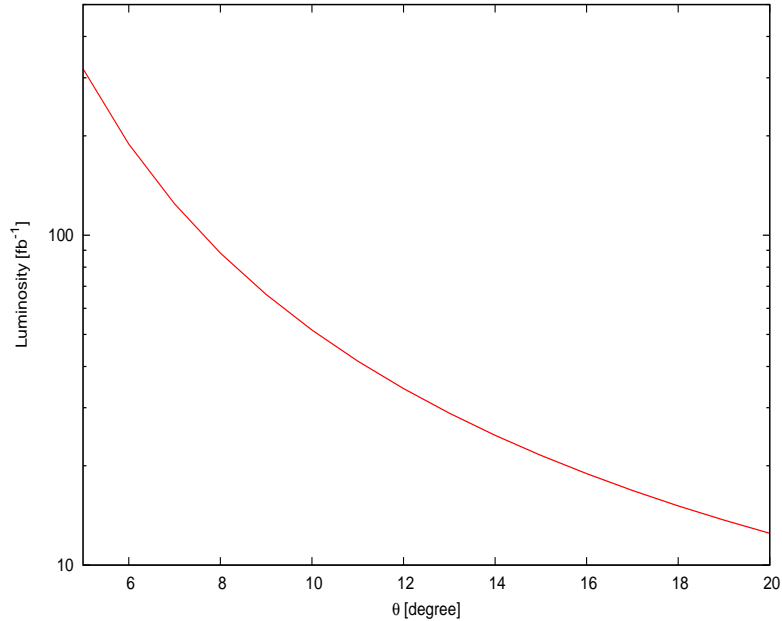}
\end{center}
\caption{ Required luminosity for 5$\sigma$ discovery as a function of the mixing angle $\theta$.}
\label{lumi}
\end{figure}    

Invisible Higgs decay is a characteristic signature of this model. The invisible decay rate directly depends on the mixing angle and hence, could be used to probe the mixing angle.  In view of Fig.~\ref{re_mass}, we impose further cuts on the $M_{Recoil}$: $123~{\rm GeV} \le M_{Recoil} \le 127~{\rm GeV}$, to enhance the signal to background ratio. In order to quantify the ability of extracting signal event,  $N_S=\sigma_S{\cal L}$, for a given integrated luminosity ${\cal L}$ over the SM background events, $N_B=\sigma_B{\cal L}$, we define the significance $S=N_S/{\sqrt {N_B+N_S}}$. In Fig.~\ref{lumi}, we have presented the required luminosity for 5$\sigma$ (i.e., $s=5$) discovery as a function of the mixing angle $\theta$ for an electron-positron collider with center-of-mass energy of 250 GeV. Fig.~\ref{lumi} shows that a 250 GeV electron-positron collider with 300 fb$^{-1}$ integrated luminosity will be able to probe $\theta$ upto $5^0$. 

\section{Summary and Conclusion}

 The fact that the Dark Matter is about  five times the ordinary  baryonic matter is well established now from several astrophysical observations. Motivated by this fact, we have presented a scenario under which the dark protons, dark neutrons and the dark nuclei belonging to the parallel Universe are taken to be the dark matter of our Universe. Assuming that the QCD scale in the dark sector ($\Lambda_{DS}$) is about five times larger than the QCD scale in the visible sector ($\Lambda_{VS}$), we get the dark matter  in the desired amount the desired amount. Assuming non-abelian Pati-Salam gauge symmetry we also get charge quantization as well as prohibit any unwanted kinetic mixings between the two photons of the respective universes. In this model the only connection between the two universes come from the respective Higgs sectors which leads to very interesting Higgs Phenomenology at the colliders. We get very interesting phenomenology in the special case when the two light Higgs bosons of the two Universes are almost degenerate in mass. Specifically, if their mass difference is  $\sim 100 $MeV , the LHC would not be able to resolve them as two separate mass peaks in its run time. We consider proposed ILC  where because of its clean environment, the precise measurements and large number of Higgs boson production, Higgs mass splittings upto $\sim 100$ MeV may be  possible with high luminosity. We investigate two scenarios where the mass difference between the two Higgs bosons are 40 MeV and 500 MeV for two mixing angles of $20^0$ and $10^0$. We find that with a $250 $ GeV ILC, for 500 MeV mass splitting we can see two clear mass peaks when the Higges decay to $b \bar{b}$ or invisibly. But for the more ambitious $\sim 40$ MeV mass splitting, it is not possible to resolve them. We also  study the sensivity to the mixing angle to the recoil mass to the $Z$-boson when the both the Higgs bosons decay invisibly to a pair of dark sector particles. We show that with a 250 GeV ILC  with 300 fb$^{-1}$ integrated luminosity it will be possible to probe the mixing angle $\theta$ between the two Higgs bosons upto $5^0$.

\noindent{\bf Acknowledgment}: During the completion of this work, SC is at the Kavli Institute of Theoretical Physics (KITP), Santa Barbara where she is a Graduate Fellow for Fall 2014. This research is supported in part by the National Science Foundation under Grant No.NSF PHY11-25915. The work of SC, KG and SN is supported in part by the U.S. Department of Energy Grant Number DE-SC0010108.



\begin{thebibliography}{999}
\bibitem{ATLAS_higgs}
ATLAS Collaboration, G. Aad et.al, Phys. Lett. B716(2012)1;
CMS Collaboration, S. Chatrchyan et. al, Phys. Lett. B716(2012)30.


\bibitem{GG}
  H. Georgi and S. L. Glashow, Phys. Rev. Lett. 32 (1974)438.
  
\bibitem{gfm}
  H. Georgi, in Proceedings of the American Institute of Physics, edited by C. E. Carlson, Meetings at College of William and Mary, 1974; H. Fritzsch and P. Minkowski, Ann. Phys. 93 (1975)193.
\bibitem{WMAP1}
D. N. Spergel \textit{et al}., Astrophys. J. Suppl. \textbf{148} (2003) 175.

G. Hainshaw \textit{et al}., Astrophys. J. Suppl. Ser. \textbf{170} (2007) 288;
L. Pagel \textit{et al}., Astrophys. J. Suppl. Ser. \textbf{170} (2007) 355;
D. N. Spergel \textit{et al}., Astrophys. J. Suppl. \textbf{170} (2007) 377;
D. Larson \textit{et al}. (WMAP Collaboration), Astrophys. J. \textbf{192} (2011) 16.
 \bibitem{dm}
  For example, see Particle Data Group Review on Dark Matter by M Drees and G. Geibier, September, 2011.

\bibitem{ww}
S. Weinberg, Phys. Rev. Lett. 40(1978)223;
F. Wilczek, Phys. Rev. Lett. 40 (1978)279.

\bibitem{kkdm}
  
  Hsin-Chia Chen, Jonathan L. Feng and Konstantin T. Matchev, Phys. Rev. Lett. 89(2002)211301;
Geraldine Servant and Tim M.P. Tait, Nucl. Phys. B650(2003)391.

\bibitem{chsw}

P. Candelas, Gary t. Horowitz, Andrew Strominger and Edward Witten, Nucl. Phys. B258(1985)46.
 
 
\bibitem{fv}

R. Foot, H. Lew and R. R. Volkas, JHEP 0007(2000)032;
H. An, S.-L. Chen, R. N. Mohapatra and Y. Zhang, JHEP 1003(2010)124.
\bibitem{plbus} 
  S.~Chakdar, K.~Ghosh and S.~Nandi,
  Phys.\ Lett.\ B {\bf 732}, 343 (2014).
\bibitem{Djouadi:2007ik} 
  G.~Aarons {\it et al.}  [ILC Collaboration],
  arXiv:0709.1893 [hep-ph].

 \bibitem{PS} 
  J.~C.~Pati and A.~Salam,
  Phys.\ Rev.\ D {\bf 10}, 275 (1974)
  [Erratum-ibid.\ D {\bf 11}, 703 (1975)].
  

  \bibitem{mm} 
  R.~E.~Marshak and R.~N.~Mohapatra,
  Phys.\ Lett.\ B {\bf 91}, 222 (1980).
\bibitem{Ritchie:1993ua} 
  J.~L.~Ritchie and S.~G.~Wojcicki,
  Rev.\ Mod.\ Phys.\  {\bf 65}, 1149 (1993),
J.~Beringer {\it et al.}  [Particle Data Group Collaboration],
  Phys.\ Rev.\ D {\bf 86}, 010001 (2012), 
  G.~Valencia and S.~Willenbrock,
  Phys.\ Rev.\ D {\bf 50}, 6843 (1994)
  [hep-ph/9409201], S.~Chakdar, T.~Li, S.~Nandi and S.~K.~Rai,
  Phys.\ Lett.\ B {\bf 718}, 121 (2012)
  [arXiv:1206.0409 [hep-ph]].


\bibitem{Senjanovic} 
  G.~Senjanovic,
  Nucl.\ Phys.\ B {\bf 153}, 334 (1979).



 \bibitem{pdg_qcd}  K.A. Olive {\it et al.} (Particle Data Group), Chin. Phys. C, {\bf 38}, 090001 (2014)


\bibitem{ILD_resolution}
  T.~Abe {\it et al.}  [ILD Concept Group - Linear Collider Collaboration],
  arXiv:1006.3396 [hep-ex];
  H.~Li {\it et al.}  [ILD Design Study Group Collaboration],
  arXiv:1202.1439 [hep-ex];
  D.~M.~Asner, T.~Barklow, C.~Calancha, K.~Fujii, N.~Graf, H.~E.~Haber, A.~Ishikawa and S.~Kanemura {\it et al.},
  arXiv:1310.0763 [hep-ph].
\bibitem{calchep}
A.~Belyaev, N.~D.~Christensen and A.~Pukhov,
  Comput.\ Phys.\ Commun.\  {\bf 184}, 1729 (2013).
\bibitem{ISR} 
  S.~Jadach and B.~F.~L.~Ward,
  Comput.\ Phys.\ Commun.\  {\bf 56}, 351 (1990);
  M.~Skrzypek and S.~Jadach,
  Z.\ Phys.\ C {\bf 49}, 577 (1991).
\bibitem{ILD} 
  T.~Behnke, J.~E.~Brau, B.~Foster, J.~Fuster, M.~Harrison, J.~M.~Paterson, M.~Peskin and M.~Stanitzki {\it et al.},
  arXiv:1306.6327 [physics.acc-ph].


\end{thebibliography}
\end{document}